\documentclass[12pt, authoryear]{elsarticle}       
\journal{arXiv.org}                                

\usepackage{amssymb,amsfonts,amsmath}
\usepackage{rotating}
\usepackage[ 
             top=1.0in, bottom=1.0in, left=1.0in, right=1.0 in]{geometry}
\usepackage[bottom]{footmisc}
\usepackage{indentfirst}
\usepackage{mdwlist,enumitem}
\usepackage{hyperref}                                            
\usepackage{epsfig,pgf}
\usepackage{rotating,longtable,booktabs, colortbl,siunitx}       
\usepackage{pdflscape}
\usepackage{threeparttable}
\usepackage[toc,page]{appendix}
\usepackage{natbib}
\usepackage{subcaption,float}
\usepackage{dcolumn}
\newcolumntype{d}[1]{D{.}{.}{#1}}
\usepackage[standard]{ntheorem}
\usepackage[algo2e]{algorithm2e}
\usepackage[T1]{fontenc}
\usepackage{lmodern}
\allowdisplaybreaks[1]                     
\theoremstyle{break}
\theorembodyfont{\normalfont}
\newtheorem{algorithm}[algocf]{Algorithm}

\begin{document}

\begin{frontmatter}
\title{\textbf{Heterogeneity in Food Expenditure amongst US families:
 Evidence from Longitudinal Quantile Regression}}
\author{Arjun Gupta}
\ead{arjung@iitk.ac.in}
\address{Department of Economic Sciences, Indian Institute of
Technology Kanpur, 208016.}

\author{Soudeh Mirghasemi}
\ead{Soudeh.Mirghasemi@hofstra.edu}
\address{Department of Economics, Hofstra University, USA}

\author{Mohammad Arshad Rahman\corref{cor1}}
\ead{marshad@iitk.ac.in}
\address{Department of Economic Sciences, Indian Institute of
Technology Kanpur, 208016.} \cortext[cor1]{Corresponding author}

\fntext[fn1]{Arjun Gupta worked on this paper while pursuing his M.S. degree
in the Department of Economic Sciences at the Indian Institute of Technology
Kanpur, India}

\begin{abstract}
Empirical studies on food expenditure are largely based on cross-section
data and for a few studies based on longitudinal (or panel) data the focus
has been on the conditional mean.  While the former, by construction,
cannot model the dependencies between observations across time, the latter
cannot look at the relationship between food expenditure and covariates
(such as income, education, etc.) at lower (or upper) quantiles, which are
of interest to policymakers. This paper analyzes expenditures on total food
(TF), food at home (FAH) and food away from home (FAFH) using mean
regression and quantile regression models for longitudinal data to examine
the impact of economic recession and various demographic, socioeconomic,
and geographic factors. The data is taken from the Panel Study of Income
Dynamics (PSID) and comprises of 2174 families in the United States (US)
observed between 2001$-$2015. Results indicate that age and education of
the head, family income, female headed family, marital status, and economic
recession are important determinants for all three types of food
expenditure. Spouse education, family size and some regional indicators are
important for expenditures on TF and FAH, but not for FAFH. Quantile
analysis reveals considerable heterogeneity in the covariate effects for
all types of food expenditure, which cannot be captured by models focused
on conditional mean. The study ends by showing that modeling conditional
dependence between observations across time for the same family unit is
crucial to reducing/avoiding heterogeneity bias and better model fitting. \\

\emph{JEL Classification}: C11, C31, C33, D10, D12, R20.
\end{abstract}

\begin{keyword}
Bayesian quantile regression, Great Recession, heterogeneity bias,
longitudinal data, mixed effects, mortgage.
\end{keyword}

\end{frontmatter}

\section{Introduction}\label{sec:Intro}


Food expenditure forms an integral part of the total family (or household)
expenditure and is often categorized into food at home (FAH), food away from
home (FAFH) and food delivered at home (FDAH). This categorization is
relevant from a health perspective and other reasons. First, the division
permits us to analyze the nutrition quality of food amongst families. This is
important because there are health implications of consuming more FAFH, as it
is considered to be less nutritious than FAH \citep{Mancino-etal-2009} and
more energy dense \citep{Binkley-2008}. Some authors have also linked more
FAFH to overweight and obesity \citep{Cai-etal-2008}. Second, the division
allows us to answer interesting policy-oriented questions. For example, what
is the effect of a female headed family on FAH expenditure or does having a
home mortgage reduce FAH and/or FAFH expenditures? Third, food assistance
programs are often designed to minimize the health risks arising from
deficient nutrition particularly amongst unemployed and lower-income groups.
This categorization can help assess the efficacy of food assistance program
on FAH expenditure of the vulnerable groups, particularly during times of
economic crisis.

\begin{sloppypar}
As a result, the study of expenditure on FAH and FAFH have attracted
considerable attention in the literature. Few previous studies using
cross-section data include \citet{Lee-Brown-1986}, \citet{Nayga-1996},
\citet{Aguiar-Hurst-2005} and \citet{Liu-et-al-2013}. \citet{Lee-Brown-1986}
employ a switching regression model on the 1977-78 Nationwide Food
Consumption Survey data to examine expenditures on FAH and FAFH amongst the
US households. \citet{Nayga-1996} utilizes the 1992 US consumer expenditure
survey (CES) data to estimate the effect of wife's education and employment
on three subcategories of food expenditure -- for prepared food, food
prepared at home, and food away from home. The modeling scheme utilized is a
generalized version of Heckman's sample selection model \citep{Heckman-1979}.
\citet{Aguiar-Hurst-2005} employs an instrumental variable linear regression
to investigate, amongst other things, the effect of anticipated (i.e.,
retirement) and unanticipated (i.e., unemployment) shock to income on TF, FAH
and FAFH expenditures. The data is taken from the Continuing Survey of Food
Intake of Individuals (CSFII, collected by the US Department of Agriculture)
and corresponds to interviews conducted between $1989-1991$ and $1994-1996$,
but the households are different in the two interviews.
\citet{Liu-et-al-2015} use a trivariate sample selection procedure to study
patterns in FAFH expenditure amongst the Chinese households. In studies such
as \citet{Liu-et-al-2013}, the use of the sample selection framework is
motivated to account for the occurrence of zero expenditures, particularly on
FAFH or in its subdivision (e.g., full-service restaurants, fast-food
restaurants and others). To get a more complete picture, readers may look
into Table~1 of \citet{Davis-2014} for a brief summary of 17 articles (out of
20) on studies related to food expenditure using cross-section data.
\end{sloppypar}

The relationship of food expenditure (at home and away from home) to other
covariates have been the focus of analysis in several cross-section studies.
They include the relation of food expenditure (of various types) to consumer
preferences \citep{Stewart-etal-2005}, family composition
\citep{Liu-et-al-2013}, race \citep{Lanfranco-etal-2002}, homeownership and
mortgages \citep{Nayga-1996, Mian-etal-2013}, wife's labor force
participation \citep{Redman-1980, Kinsey-1983, Darian-Klein-1989, Yen-1993,
Nayga-1996}, children's welfare \citep{Handa-1996}, and obesity
\citep{Drichoutis-etal-2012}. Some authors have also examined the effects of
tax on food expenditure. For example, \citet{Zheng-etal-2019} examines the
impact of tax on expenditure in grocery food (i.e., FAH) and restaurant food
(i.e., FAFH) using a weekly data observed between April 2012$-$January 2013,
collected by United States Department of Agriculture (USDA). They find that
tax on grocery (restaurant food) reduces expenditure on grocery (restaurant
food) and increases expenditure on restaurant food (grocery).

The above paragraphs clearly indicate that there are ample cross-section
studies on food expenditure, but panel or longitudinal studies are rather
lacking with few exceptions. \citet{Cai-etal-2008} presents a state-level
analysis of different types of food expenditure on overweight rates, obesity
rates and combined rates (the sum of overweight and obesity rates) using data
from the Behavioral Risk Factor Surveillance System. The primary finding is
that FAH (FAFH) expenditure is negatively (positively) associated to obesity
and combined rates, and both FAH and FAFH expenditures do not significantly
affect overweight rates. The only panel study mentioned in \citet{Davis-2014}
is the article by \citet{Gelber-Mitchell-2012}, where they use PSID and time
diary data between $1975-2004$, and find that for a decrease in income tax
(i.e., incentive to join the labor force increases) single women are much
more likely to increase FAFH expenditure to substitute for housework compared
to single men. At the same time, the effect on FAH expenditure is
statistically insignificant. \citet{Kohara-Kamiya-2016} use a panel data on
Japanese households for the period 2004$-$2006 and find that mothers' labor
supply decision has a negative effect on food produced at home. Moreover, the
negative effect is common for all economic classes and more pronounced for
the low economic class. Besides, there are abundant studies that examine the
impact on food expenditure from participating in Supplemental Nutrition
Assistance Program (SNAP), formerly known as Food Stamp Program (FSP)
\footnote{The American Recovery and Reinvestment Act (ARRA) of 2009, renamed
the FSP to SNAP and increased benefits by an average of \$80 per household.
However, a common variable to capture SNAP participation pre- and post-ARRA
is not available in PSID.}. Few articles from this literature\footnote{Within
the SNAP literature, the central debate is whether households respond
similarly to an increase in cash income and in-kind transfer (food coupons).
While some researchers, such as \citet{Hoynes-Schanzenbach-2009}, have found
that the response is similar; others such as \citet{Beatty-Tuttle-2014} have
found that households increase in food expenditure is more when given an
in-kind transfer (food stamps) as compared to cash income.} includes
\citet{Hoynes-Schanzenbach-2009}, \citet{Wilde-etal-2009},
\citet{Beatty-Tuttle-2014} and \citet{Burney-2018}. However, these studies
focus on the conditional mean of the response variable and thus cannot
explain the relationship at the quantiles.

The current study takes a broader perspective and looks at expenditures on
total food (TF), food at home (FAH), and food away from home (FAFH), and
explains its variation based on various demographic, socioeconomic and
geographic factors including mortgage and recession. The data is taken from
Panel Study of Income Dynamics (PSID) and is composed of 2174 family units
observed over the period $2001-2015$. Since ours is a panel data, we exploit
a longitudinal or panel regression framework that can accommodate both common
(fixed-effects) and individual-specific (random-effects) parameters (hence
also known as mixed effects model in Statistics)\footnote{The terms
fixed-effects and random-effects have been used to mean different things in
the literature and there is no agreed-upon definition. In this paper,
fixed-effects refers to regression coefficients that do not differ across $i$
(or individuals) and random-effects mean regression coefficients that differ
across $i$ \citep[see][Ch.~10]{Greenberg-2012}. Andrew Gelman lists five
different definitions of fixed-effects and random-effects at
{\url{https://statmodeling.stat.columbia.edu/2005/01/25/why_i_dont_use/}}.
But again, there are other popular definitions such as in Classical
econometrics where fixed-effects means that the unobserved
individual-specific heterogeneity are correlated with the regressors, while
random-effects imply zero correlation (or more strongly statistical
independence) between individual-specific heterogeneity and the regressors
\citep[see][]{Cameron-Trivedi-2005,Wooldridge-2010,Hsiao-2014,Greene-2017}.}.
However, mean longitudinal regression is not capable of capturing the
heterogeneity in covariate effects across the conditional distribution of the
response variable. To overcome this limitation, we study the heterogeneous
effect of the covariates on food expenditure (TF, FAH and FAFH) using a
quantile model for longitudinal data that accommodates both common effects
and individual-specific effects, also known as quantile mixed models.

This paper contributes to the literature in at least three different ways.
First, quantile longitudinal regression provides a comprehensive
understanding of food expenditure pattern of family units to variation in
covariates by providing estimates at different quantiles. The method is
robust compared to standard longitudinal models where the focus is on the
mean, because amongst other things it is unaffected by the presence of
outliers in the data. Second, this study adds to the understanding of the
differences in food expenditure pre-, during- and after- the Great Recession.
This enables us to capture patterns linking recession and food expenditure by
categories which we explore in this study. To our knowledge, this is the
first attempt to examine the effects of the Great Recession on food
expenditure at home and away from home within a quantile panel data
framework. Third, longitudinal data allows us to model the behavior of family
units over time, which provides an advantage to control for unobserved
heterogeneity leading to more robust estimates. As shown in this paper, it is
important to control for this repeated behavior because models which treat
unobserved heterogeneity as a part of error term often result in inconsistent
estimates and may lead to incorrect policy inference.

The remaining paper is organized as follows. Section~\ref{sec:Methodology}
lays out the basic framework of the mean regression and quantile regression
models for longitudinal data that we employ in our analysis.
Section~\ref{sec:Data} presents a descriptive summary of the data and
discusses the trends in variables over the time period of our study.
Section~\ref{sec:Results} presents the results from the aforementioned
regression models and shows the consequences of not modeling
individual-specific heterogeneity. Finally, Section~\ref{sec:Conclusion}
presents concluding remarks.

\section{Methodology}\label{sec:Methodology}

This section presents the mean regression for longitudinal data model and
outlines the Bayesian approach for its estimation \citep{Chib-1999,
Greenberg-2012}. Thereafter, we present the Bayesian quantile regression for
longitudinal data model and its estimation algorithm, which is inspired from
\citet{Luo-2012} and \citet{Rahman-Vossmeyer-2019}.


\subsection{Mean Regression for Longitudinal Data}\label{subsec:MRLD}

The longitudinal data model can be expressed in terms of the following
equation,
\begin{align}
y_{it} = x'_{it}\beta + s'_{it}\alpha_{i} + \epsilon_{it}, \hspace*{1cm} \forall
\hspace*{0.2cm} i = 1,\ldots,n, \hspace*{0.5cm} t = 1,\ldots,T,
\label{eq:mrld1}
\end{align}
where $y_{it}$ denotes the value of the response $y$ for the $i$-th
individual at the $t$-th time period, $x'_{it}$ is a $1 \times k$ vector of
explanatory variables, $\beta$ is $k \times 1$ vector of common
(fixed-effects) parameters, $s'_{it}$ is a $1 \times l$ vector of covariates
(often a subset of $x_{it}$) with individual-specific effects, $\alpha_{i}$
is an $l \times 1$ vector of individual-specific (random-effects) parameters
included to capture the marginal dependence between observations on the same
individual, and $\epsilon_{it}$ is the error term assumed to independently
and identically distributed (\emph{iid}) as a normal distribution i.e.,
$\epsilon_{it} \overset{iid}{\sim} N(0, h^{-1})$ for all values of
$i=1,\ldots,n$; $t=1,\cdots,T$, where $h^{-1}$ is the variance. The
distributional assumption on the error implies that $y_{it}$, conditional on
$\alpha_{i}$ are independently distributed as a normal distribution i.e.,
$y_{it}|\alpha_{i} \sim N(x'_{it}\beta + s'_{it} \alpha_{i},h^{-1})$ for all
$i=1,\ldots,n$; $t=1,\cdots,T$.

In this paper, the response variable $y$ will either be TF, FAH or FAFH
expenditures. The vector $x_{it}$ will consist of a common intercept and a
host of covariates related to demographic, socioeconomic and geographic
factors. Lastly, the vector of covariates with individual-specific effects
$s'_{it}$ will consist of an intercept and inverse-hyperbolic sine
transformation of income.

To proceed with the Bayesian estimation of the longitudinal model, we first
stack the model for each individual $i$. This is convenient for multiple
reasons including reducing the computational burden. We define $y_{i} =
(y_{i1},\ldots,y_{iT})'$, $X_{i} = (x'_{i1}, x'_{i2},\ldots,x'_{iT})'$,
$S_{i} = (s'_{i1}, s'_{i2}, \ldots,s'_{iT})'$, $\epsilon_{i} =
(\epsilon_{i1}, \ldots,\epsilon_{iT})'$. The resulting stacked model can be
written as,
\begin{equation}
\begin{split}
y_{i} & =  X_{i} \beta  + S_{i}\alpha_{i} + \epsilon_{i}, \hspace*{1cm}
\textrm{for} \; i=1,\cdots,n,
\\
& \alpha_{i}|\Sigma  \sim N_{l}(0, \Sigma),
\\
\beta &  \sim N_{k}(\beta_{0}, B_{0}), \qquad
\Sigma^{-1} \sim Wish(\nu_{0}, D_{0}), \qquad h \sim Ga(c_{0}/2, d_{0}/2),
\end{split}
\label{eq:mrld2}
\end{equation}
where we assume that $\alpha_{i}|\Sigma$ are mutually independent and
identically distributed as $N_{l}(0, \Sigma)$, and the last line represents
the prior distributions, with $N$, $Wish$ and $Ga$ denoting the normal,
Wishart and gamma distributions, respectively. The model given by
equation~\eqref{eq:mrld2} implies that the conditional density
$y_{i}|\alpha_{i} \sim N( X_{i} \beta  + S_{i}\alpha_{i}, h^{-1}I_{T})$ for
$i=1,\ldots,n$. The complete data density is then given by,
\begin{equation*}
f(y,\alpha|\beta,h,\Sigma) = \prod_{i=1}^{n}
f(y_{i},\alpha_{i}|\beta,h,\Sigma) = \prod_{i=1}^{n}
f(y_{i}|\beta,\alpha_{i},h) \pi(\alpha_{i}|\Sigma),
\end{equation*}
which is equivalent to the complete data likelihood when viewed as a function
of the parameters.

\begin{table*}[b!]
\begin{algorithm}
\label{algo1}
\rule{\textwidth}{0.5pt}\small{
\begin{enumerate}[itemsep=-2ex]
\item  Sample $(\beta, \alpha)$ in one block as follows:
\begin{enumerate}[leftmargin=3ex]
\item  Let $\Psi_{i} = S_{i}\Sigma S_{i}'+ h^{-1}I_{T}$.
       Sample $\beta$ marginally of $\alpha$ from
       $\beta|y,h,\Sigma $ $\sim$  $N\big(\widetilde{\beta}, \widetilde{B}\big)$, where,
       \begin{equation*}
       \widetilde{B}^{-1} = \bigg(\sum_{i=1}^{n}
       X'_{i} \Psi_{i}^{-1}   X_{i}
       + B_{0}^{-1} \bigg), \quad
       \mathrm{and} \quad
       \widetilde{\beta} = \widetilde{B}\left(\sum\limits_{i=1}^{n}X_{i}'\Psi_{i}^{-1}
                          y_{i}+ B_{0}^{-1}\beta_{0} \right).
       \end{equation*}
\item  Sample $\alpha_{i}|y,\beta,h,\Sigma$ $\sim$  $N\big(\widetilde{a},
       \widetilde{A}\big)$
       for $i=1,\cdots,n$, where,
       \begin{equation*}
       \widetilde{A}^{-1} = \left(hS_{i}'S_{i} + \Sigma^{-1}\right),
       \quad \mathrm{and} \quad
       \tilde{a} = \widetilde{A} \Big(h S_{i}'\big(y_{i}-X_{i}\beta\big) \Big).
       \end{equation*}
\end{enumerate}
\item  Sample $\Sigma^{-1}|\alpha \sim Wish\big(\nu_{1},D_{1}\big)$, where
       $\nu_{1} = (\nu_{0} + n)$, and
       $D_{1}^{-1}   = \Big( \displaystyle D_{0}^{-1}+
       \sum\limits_{i=1}^{n}\alpha_{i}\alpha_{i}' \Big)$.\\
\item  Sample $h|y,\beta,\alpha$ $\sim$
       $Ga\big(c_{1}/2,d_{1}/2 \big)$ where,
       \begin{equation*}
       c_{1}  = \left(c_{0} + nT\right),
       \quad \mathrm{and} \quad
       d_{1}  = d_{0} + \sum_{i=1}^{n}(y_{i}-X_{i}
       \beta-S_{i}{\alpha}_{i})'(y_{i}-X_{i}\beta-S_{i}{\alpha}_{i}).
       \end{equation*}
\end{enumerate}}
\rule{\textwidth}{0.5pt}
\end{algorithm}
\end{table*}

By Bayes' theorem, the complete data posterior density can be written as
product of the complete data likelihood times the prior distributions as
follows,
\allowdisplaybreaks{
\begin{equation}
\begin{split}
& \pi(\beta, \alpha, \Sigma^{-1},h|y)  \propto  \Big\{\prod_{i=1}^{n}
f(y_{i}|\beta,\alpha_{i},h) \pi(\alpha_{i}|\Sigma) \Big\}
\pi(\beta) \pi(\Sigma^{-1}) \pi(h)\\
& \quad \propto h^{nT/2}\exp\bigg[-\frac{h}{2}
\sum\limits_{i=1}^{n}(y_{i}-X_{i}\beta-S_{i}
{\alpha}_{i})'(y_{i}-X_{i}\beta-S_{i}{\alpha}_{i})\bigg]\\
& \qquad \times |\Sigma|^{-\frac{n}{2}}
\exp\bigg[-\frac{1}{2}\sum\limits_{i=1}^{n}\alpha_{i}'\Sigma^{-1}\alpha_{i}\bigg]
\exp\Big[-\frac{1}{2}(\beta - \beta_{0})'B_{0}^{-1}(\beta - \beta_{0})\Big] \\
& \qquad \times |\Sigma^{-1}|^\frac{(\nu_{0}-l-1)}{2}
\exp\bigg[-\frac{1}{2}tr(D_{0}^{-1}\Sigma^{-1})\bigg] \times
h^{\frac{c_{0}}{2}-1} \exp\bigg[-\frac{d_{0}h}{2}\bigg].
\end{split}
\label{eq:mrldPost}
\end{equation}
}

The conditional posterior distributions are derived from the complete data
posterior (Equation~\ref{eq:mrldPost}) and the model is estimated using Gibbs
sampling, a well known Markov chain Monte Carlo method
\citep{Geman-Geman-1984,Casella-George-1992}. The MCMC algorithm for
estimating the model is presented in Algorithm~\ref{algo1}. The parameters
$(\beta,\alpha)$ are sampled jointly to avoid correlation between the
parameters, because the covariates in $s_{it}$ are often a subset of $x_{it}$
\citep[][Chap. 10]{Greenberg-2012}. Specifically, we first sample $\beta$
(marginally of $\alpha$, but conditional on other model parameters) from an
updated normal distribution and then sampled $\alpha$ (conditional on $\beta$
and other model parameters) from its updated normal distribution. The
precision matrix $\Sigma^{-1}$ is sampled from an updated Wishart
distribution and finally, the precision parameter $h$ is sampled from an
updated gamma distribution.


\subsection{Quantile Regression for Longitudinal Data}\label{subsec:QRLD}

The quantile regression for longitudinal data can be expressed in terms of
the following equation,
\begin{align}
y_{it} = x'_{it}\beta + s'_{it}\alpha_{i} + \epsilon_{it}, \hspace*{1cm} \forall
\hspace*{0.2cm} i = 1,\ldots,n, \hspace*{0.5cm} t = 1,\ldots,T,
\label{eq:qrld1}
\end{align}
where all the notations are same as in Section~\ref{subsec:MRLD}, except that
the errors are assumed to be \emph{i.i.d.} as an asymmetric Laplace (AL)
distribution, i.e., $\epsilon_{it} \overset{iid}{\sim} AL(0,h^{-1},p)$, where
$h^{-1}$ is the inverse of the scale parameter and $p$ denotes a quantile.
This implies that $y_{it}$, conditional on $\alpha_{i}$, are independently
distributed as an AL distribution i.e., $y_{it}|\alpha_{i} \sim AL(x'_{it}
\beta + s'_{it}\alpha_{i}, h^{-1},p)$ for $i=1,\cdots,n$, $t=1,\ldots,T$.
Note that the error distribution is assumed to be AL to form a working
likelihood because the quantile loss function appears in the exponent of an
AL distribution \citep[see][]{Yu-Moyeed-2001,Rahman-2016}. The resulting
conditional quantile function for response $y_{it}$ is,
\begin{align*}
Q_{y_{it}}(p|x_{it},\alpha_{i}) = x'_{it} \beta + s'_{it}\alpha_{i},
\end{align*}
where $Q_{y_{it}} \equiv F_{y_{it}}^{-1}(\cdot)$ is the inverse of the
cumulative distribution function of the outcome variable conditional on the
individual specific parameters and the covariates.

We can directly work with the AL distribution, however, it is not convenient
for Gibbs sampling. So, as proposed in \citet{Kozumi-Kobayashi-2011}, we make
use of the normal-exponential mixture representation of the AL distribution,
\begin{equation}
\epsilon_{it} = h^{-1} \theta w_{it} + h^{-1} \tau \sqrt{w_{it}} \, u_{it},
\hspace{0.75in}
\forall \; i=1,\ldots,n; \; t = 1, \ldots, T,
\label{eq:normal-exp}
\end{equation}
where $u_{it} \sim N(0,1)$ is mutually independent of $w_{it}
\sim\mathcal{E}(1)$, $\theta = \frac{1-2p}{p(1-p)}$, $\tau =
\sqrt{\frac{2}{p(1-p)}}$, and the symbol $\mathcal{E}$ denotes an exponential
distribution. The resulting quantile regression for longitudinal data model
can be expressed as,
\begin{align}
y_{it} = x'_{it}\beta + s'_{it}\alpha_{i} +  \theta \nu_{it} +
\tau \sqrt{h^{-1} \nu_{it}} \, u_{it}, \hspace*{1cm} \forall
\hspace*{0.2cm} i = 1,\ldots,n, \hspace*{0.5cm} t = 1,\ldots,T.
\label{eq:qrld2}
\end{align}
where we have used the transformation $\nu_{it} = w_{it}/h$, since the
presence of the scale parameter in the conditional mean is not conducive to
Gibbs sampling \citep{Kozumi-Kobayashi-2011,Rahman-Karnawat-2019}. See also
\citet{Bresson-etal-2020} and \citet{Ojha-Rahman-2020}, where the scale is
fixed at 1 to identify the parameters of quantile regression with binary
outcomes.

To proceed with the Bayesian estimation, we again stack the model across $i$
for reasons mentioned earlier. Define $y_{i} = (y_{i1}, \ldots ,y_{iT})'$,
$X_{i} = (x'_{i1}, x'_{i2},\ldots,x'_{iT})'$, $S_{i} = (s'_{i1},
s'_{i2},\ldots,s'_{iT})'$, $D_{\tau\sqrt{\frac{\nu_{i}}{h}}}  =
diag(\tau\sqrt{\frac{\nu_{i1}}{h}}, \ldots ,\tau\sqrt{\frac{\nu_{iT}}{h}})$,
$u_{i} = (u_{i1}, \ldots ,u_{iT})'$, and lastly $\nu_{i} = (\nu_{i1}, \ldots
,\nu_{iT})'$. The resulting stacked quantile regression for longitudinal data
can be written as,
\begin{equation}
\begin{split}
y_{i}  & =  X_{i} \beta  + S_{i}\alpha_{i} + \theta \nu_{i} +
D_{\tau \sqrt{\frac{\nu_{i}}{h}}} \; u_{i}, \hspace*{1cm} \textrm{for} \; i=1,\ldots,n,
\\
\alpha_{i}|\Sigma & \sim N_{l}(0, \Sigma), \hspace{0.6in}
\nu_{it} \sim \mathcal{E}(1/h),
\hspace{0.7in} u_{it} \sim N(0,1),
\\
\beta & \sim N_{k}(\beta_{0}, B_{0}), \qquad \Sigma^{-1} \sim Wish(\nu_{0},
D_{0}), \qquad h \sim Ga(c_{0}/2, d_{0}/2),
\end{split}
\label{eq:qrld3}
\end{equation}
where we assume $\alpha_{i}|\Sigma$ are mutually independent and identically
distributed as $N_{l}(0, \Sigma)$, and the last line represents the prior
distributions of the model parameters. The quantile model given by
Equation~\eqref{eq:qrld3} implies that the conditional density
$y_{i}|\alpha_{i} \sim N(X_{i}\beta + S_{i} \alpha_{i} + \theta \nu_{i},
D^{2}_{\tau \sqrt{ \frac{\nu_{i}}{h}}})$ for $i=1,\ldots,n$. The complete
data density is then given by $ f(y,\alpha|\beta,v,h,\Sigma) = \displaystyle
\prod_{i=1}^{n} f(y_{i},\alpha_{i}|\beta,\nu_{i},h,\Sigma) = \displaystyle
\prod_{i=1}^{n} f(y_{i}|\beta,\alpha_{i},\nu_{i},h) \pi(\alpha_{i}|\Sigma)$.

\begin{table*}[b!]
\begin{algorithm}
\label{algo2} \rule{\textwidth}{0.5pt} \small{
\begin{enumerate}[itemsep=-2ex]
\item  Sample $(\beta, \alpha)$ in one block as follows:
\begin{enumerate}[leftmargin=3ex]
\item  Let $\Omega_{i} =
       \left(S_{i}\Sigma S'_{i} + D_{\tau \sqrt{\frac{\nu_{i}}{h}}}^{2}\right)$.
       Sample $\beta$ marginally of $\alpha$ from
       $\beta|y,\nu,\Sigma,h$ $\sim$  $N\big(\widetilde{\beta},
       \widetilde{B}\big)$, where,
       \begin{equation*}
       \widetilde{B}^{-1} = \bigg(\sum_{i=1}^{n}
       X'_{i} \Omega_{i}^{-1}   X_{i}
       + B_{0}^{-1} \bigg), \quad
       \mathrm{and} \quad
       \widetilde{\beta} =
       \widetilde{B}\left(\sum\limits_{i=1}^{n}X_{i}'\Omega_{i}^{-1}
       (y_{i}-\theta \nu_{i})+ B_{0}^{-1}\beta_{0} \right).
       \end{equation*}
\item  Sample $\alpha_{i}|y,\beta,\nu,h,\Sigma$ $\sim$  $N\big(\widetilde{a},
       \widetilde{A}\big)$
       for $i=1,\ldots,n$, where,
       \begin{equation*}
       \widetilde{A}^{-1} = \left(S'_{i} \, D^{-2}_{\tau \sqrt{\frac{\nu_{i}}{h}}}
       \, S_{i} + \Sigma^{-1}\right),
       \quad \mathrm{and} \quad
       \widetilde{a} = \widetilde{A} \left(S'_{i} D^{-2}_{\tau \sqrt{\frac{\nu_{i}}{h}}} \,
       \big(y_{i} - X_{i} \beta - \theta \nu_{i} \big)   \right).
       \end{equation*}
\end{enumerate}
\item  Sample $\nu_{it}|y_{it},\beta,\alpha_{i},h$ $\sim$  $GIG \,
       \big(0.5, \widetilde{\lambda}_{it}, \widetilde{\eta}\big)$ for
       $i=1,\ldots,n$ and $t=1,\ldots,T$, where,
       \begin{equation*}
       \widetilde{\lambda}_{it} = h\bigg( \frac{ y_{it} - x'_{it}\beta - s'_{it}
       \alpha_{i}}{\tau} \bigg)^{2} \quad \mathrm{and} \quad \widetilde{\eta}
       = h \bigg(\frac{\theta^{2}}{\tau^{2}} + 2 \bigg).
       \end{equation*}
\item  Sample $\Sigma^{-1}|\alpha \sim Wish\big(\nu_{1},D_{1}\big)$, where
       $\nu_{1} = (\nu_{0} + n)$, and
       $D_{1}^{-1} = \Big( \displaystyle D_{0}^{-1}+\sum\limits_{i=1}^{n}
       \alpha_{i}\alpha_{i}' \Big)$.\\
\item  Sample $h|y,\beta,\alpha, \nu \sim Ga\Big(c_{1}/2,d_{1}/2\Big)$ where,
       \begin{equation*}
       c_{1}  = \left(c_{0}+3nT\right),
       \quad \mathrm{and} \quad
       d_{1}  = d_{0}+2\sum\limits_{i=1}^{n}\sum
       \limits_{t=1}^{T}v_{it}+\sum\limits_{i=1}^{n}\sum\limits_{t=1}^{T}
       \frac{\big(y_{it} - x_{it}'\beta-s_{it}'\alpha_{i}-\theta
       \nu_{it}\big)^{2}}{\tau^{2} \nu_{it}}.
       \end{equation*}
\end{enumerate}}
\rule{\textwidth}{0.5pt}
\end{algorithm}
\end{table*}

Once again, we employ the Bayes' theorem to obtain the complete data
posterior as the product of the complete data likelihood times the prior
distributions as follows:
\begin{equation}
\begin{split}
& \pi(\beta,\alpha,\nu,\Sigma^{-1},h|y) \propto
\bigg\{\prod\limits_{i=1}^{n} f(y_{i}|\beta,\alpha_{i},\nu_{i},h)
\pi(\alpha_{i}|\Sigma)\pi(\nu_{i})\bigg\} \pi(\beta) \pi(\Sigma^{-1}) \pi(h)\\
& \propto \prod_{i=1}^{n}  \bigg\{ |D^{2}_{\tau\sqrt{\frac{\nu_{i}}{h}}}|^{-\frac{1}{2}}
\exp\bigg[-\frac{1}{2} (y_{i}-X_{i}\beta-S_{i}{\alpha}_{i}-\theta \nu_{i})'
D^{-2}_{\tau\sqrt{\frac{v_{i}}{h}}}
(y_{i}-X_{i}\beta-S_{i}{\alpha}_{i}-\theta \nu_{i})\bigg] \bigg\} \\
& \quad \times |\Sigma^{-1}|^{\frac{n}{2}}\exp\bigg[-\frac{1}{2}
\sum\limits_{i=1}^{n}\alpha_{i}'\Sigma^{-1}\alpha_{i}\bigg]
\times  h^{nT} \exp\bigg[-h\sum\limits_{i=1}^{n}
\sum\limits_{t=1}^{T} \nu_{it} \bigg]
\times h^{\frac{c_{0}}{2}-1}
\exp\Big(-\frac{d_{0}h}{2}\Big)\\
& \quad \times \exp\bigg[-\frac{1}{2}(\beta - \beta_{0})'B_{0}^{-1}
(\beta - \beta_{0})\bigg] \times |\Sigma^{-1}|^\frac{(\nu_{0}-l-1)}{2}
\exp\Big[-\frac{1}{2}tr(D_{0}^{-1}\Sigma^{-1})\Big].
\label{eq:qrldPost}
\end{split}
\end{equation}

\begin{sloppypar}
The conditional posteriors can be derived from the joint posterior
distribution (Equation~\ref{eq:qrldPost}) and the model can be estimated
using Gibbs sampling as presented in Algorithm~\ref{algo2}. Specifically, we
sample $\beta$ and $\alpha$ in a single block to elude the problem of poor
mixing due to correlation between the parameters for reasons mentioned
earlier \citep[see also][]{Rahman-Vossmeyer-2019, Bresson-etal-2020}. The
common effects parameters $\beta$, marginally of $\alpha$, are sampled from
an updated normal distribution and the individual-specific parameters
$\alpha_{i}$'s are sampled from their respective updated normal distribution.
The mixture variable $\nu$ is sampled component-wise from an updated
generalized inverse Gaussian (GIG) distribution \citep{Devroye-2014}. The
precision matrix $\Sigma^{-1}$ is sampled from an updated Wishart
distribution and the parameter $h$ is sampled from an updated gamma
distribution.
\end{sloppypar}

\section{Data}\label{sec:Data}

The current study utilizes data from the Panel Study of Income Dynamics
(PSID), which began in 1968 and is the longest running longitudinal household
survey in the world. We constructed a balanced panel of 2174 family units
with data for each alternate year, i.e., 2001, 2003, 2005, 2007, 2009, 2011,
2013 and 2015. This is because beginning 1997, the PSID collects data every
alternate year. Our constructed data has information on different types of
food expenditures, considered as dependent variables, and a host of
socioeconomic, demographic and geographic variables which are used as
covariates or independent variables in our study. Table
\ref{Table:DataSummary} presents a descriptive summary of the variables
considered in our analysis.

The primary variable of interest is the food expenditure of a family unit,
which the PSID categorizes into three types: food at home (FAH), food away
from home (FAFH) and food delivered at home (FDAH). The sum of these three
expenditures yield total food (TF) expenditure of the family unit. The
variable FAH represents the annualized expenditure of family unit at home and
in our sample lies between \$0 and \$36400. There are only few observations
with zero value for FAH. Similarly, the variable FAFH represents annualized
food expenditure away from home and in the sample lies in the range \$0 to
\$44,200. The zero values for FAFH is small at 5.7\% of the total number of
observations. All observations with zero TF expenditure were removed from the
sample. Our study considers expenditure on TF, FAH and FAFH as the dependent
variable in different regressions. The expenditure on FDAH is dropped due to
large number of zero values, which makes censoring important and a sample
selection framework more appropriate.

\begin{table}[t]
\footnotesize \def\sym#1{\ifmmode^{#1}\else\(^{#1}\)\fi}
\setlength{\extrarowheight}{1.2pt}
\begin{center}
\begin{tabular}{l*{8}{c}}
\toprule
\multicolumn{1}{c}{Variables$\backslash$Years} &\multicolumn{1}{c}{2001}
&\multicolumn{1}{c}{2003}  &\multicolumn{1}{c}{2005}
&\multicolumn{1}{c}{2007}  &\multicolumn{1}{c}{2009}
&\multicolumn{1}{c}{2011}  &\multicolumn{1}{c}{2013}    &\multicolumn{1}{c}{2015}\\
\hline     \\
TF/1000
&    6.70  &   6.89   &   7.40   &   7.93  &   7.91  &   8.22  &   8.56   &   8.90  \\
&   (3.62) &  (3.73)  &  (4.18)  &  (4.64) &  (4.52) &  (4.75) &  (5.18)  &  (5.50)  \\
FAH/1000
&    4.60  &   4.70   &   5.00   &   5.42  &   5.57  &   5.79  &   6.02   &   6.17  \\
&   (2.64) &  (2.64)  &  (2.82)  &  (3.17) &  (3.26) &  (3.44) &  (3.71)  &  (3.79)  \\
FAFH/1000
&    1.99  &   2.04   &   2.28   &   2.40  &   2.24  &   2.33  &   2.44   &   2.63  \\
&   (1.96) &  (2.02)  &  (2.48)  &  (2.65) &  (2.32) &  (2.43) &  (2.65)  &  (2.85)  \\

Head Age
&   44.96  &  46.97   &  48.93   &  50.96  &  52.95  &  54.96  &  56.95   &  58.98  \\
&  (12.36) & (12.37)  & (12.37)  & (12.37) & (12.36) & (12.37) & (12.36)  & (12.33) \\
Head Edu
&   13.38  &  13.43   &  13.43   &  13.43  &  13.66  &  13.66  &  13.68   &  13.69  \\
&   (2.68) &  (2.64)  &  (2.64)  &  (2.64) &  (2.62) &  (2.62) &  (2.63)  &  (2.63) \\
Spouse Edu
&    9.50  &   9.63   &   9.78   &   9.90  &  10.16  &  10.09  &   9.99   &   9.93  \\
&   (6.47) &  (6.44)  &  (6.40)  &  (6.36) &  (6.51) &  (6.53) &  (6.61)  &  (6.66) \\

Family Size
&    2.94  &   2.91   &   2.87   &   2.82  &   2.76  &   2.66  &   2.58   &   2.48  \\
&   (1.47) &  (1.45)  &  (1.42)  &  (1.43) &  (1.43) &  (1.39) &  (1.37)  &  (1.32) \\
Family Income/10000
&    7.72  &   7.97   &   8.78   &   9.22  &   9.59  &   9.34  &   9.83   &   9.94  \\
&   (8.46) & (12.32)  & (16.05)  &  (9.48) &  (9.74) & (10.31) & (13.23)  & (10.03)  \\


Head Emp
&    0.85  &   0.85   &   0.84   &   0.81  &   0.73  &   0.69  &   0.66   &   0.62  \\
Head Female
&    0.18  &   0.18   &   0.18   &   0.18  &   0.18  &   0.18  &   0.18   &   0.18  \\
Married
&    0.68  &   0.69   &   0.70   &   0.71  &   0.71  &   0.71  &   0.70   &   0.70  \\

Single
&    0.15  &   0.13   &   0.11   &   0.11  &   0.10  &   0.10  &   0.10   &   0.09  \\
Homeowner
&    0.75  &   0.78   &   0.80   &   0.81  &   0.81  &   0.81  &   0.81   &   0.81  \\
Mortgage
&    0.59  &   0.60   &   0.61   &   0.61  &   0.61  &   0.57  &   0.55   &   0.53  \\

White
&    0.68  &   0.68   &   0.71   &   0.71   &   0.71  &   0.71  &   0.71   &   0.71  \\
Non-White
&    0.32  &   0.32   &   0.29   &   0.29  &   0.29  &   0.29  &   0.29   &   0.29  \\
Recession
&    1.00  &   0.00   &   0.00   &   1.00  &   1.00  &   0.00  &   0.00   &   0.00  \\

Northeast
&    0.16  &   0.16   &   0.16   &   0.16  &   0.16  &   0.16  &   0.16   &   0.15  \\
West
&    0.20  &   0.19   &   0.19   &   0.20  &   0.19  &   0.19  &   0.19   &   0.19  \\
South
&    0.38  &   0.38   &   0.38   &   0.38  &   0.39  &   0.39  &   0.39   &   0.40  \\

\bottomrule
\end{tabular}
\caption{Data Summary - The table presents the mean and standard deviation
(in parenthesis) of the continuous variables and proportion
of the categorical variables for each considered year. \label{Table:DataSummary}}
\end{center}
\end{table}

\begin{figure}[!t]
	\centerline{
		\mbox{\includegraphics[width=6.75in, height=6.5in]{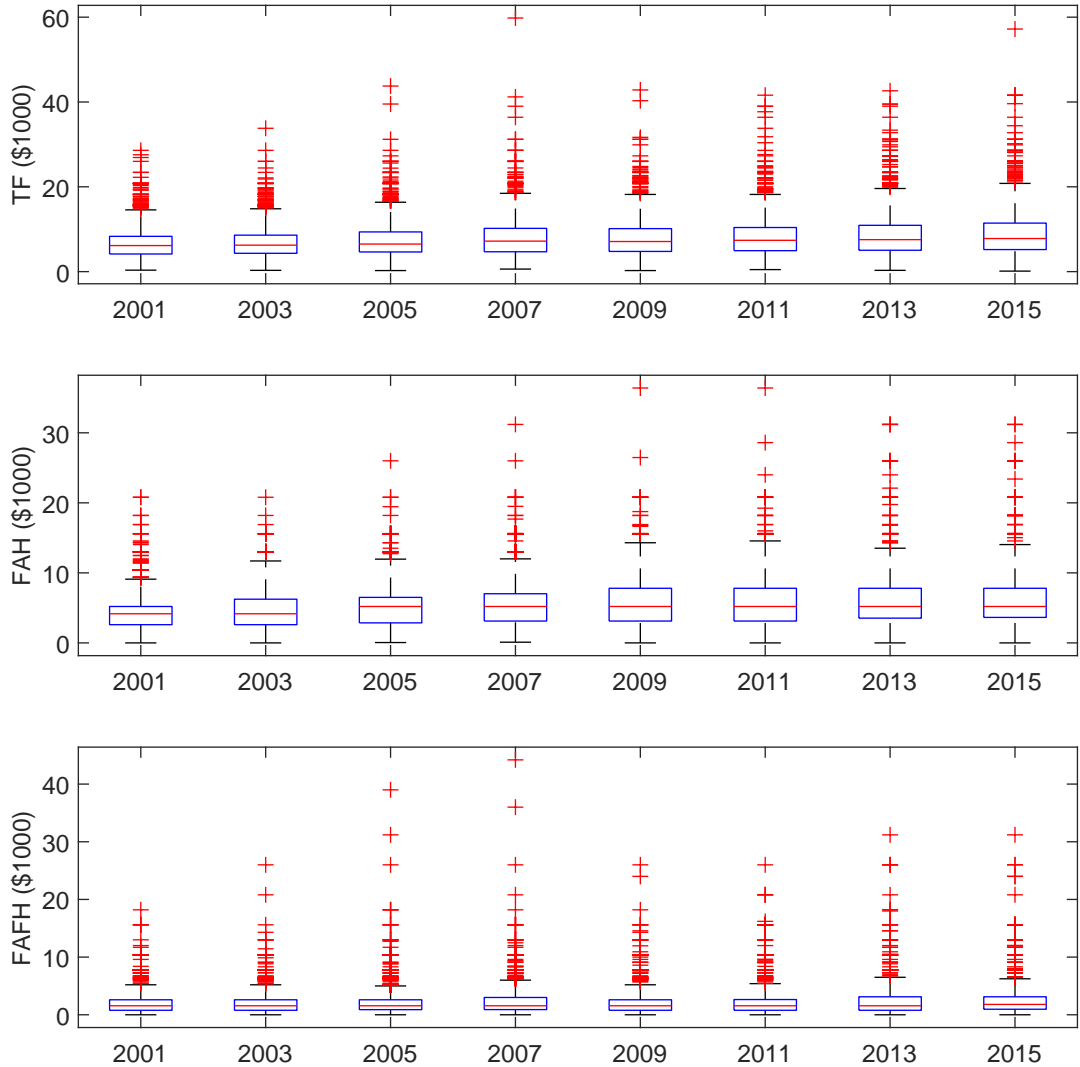}} }
\vspace{-2pc}
	\caption{Box plot for different types of food expenditure.}
\label{fig:boxplot}
\end{figure}

An interesting characteristic about the distribution of food expenditures is
that they are positively skewed. Figure~\ref{fig:boxplot} presents a box plot
of the different types of food expenditure utilized in the study. Each box
plot represents the distribution of food expenditure for a particular year.
In each box plot, the solid line within the box shows the median value, while
the bottom and top of the box represent the 25th and 75th percentiles,
respectively. The vertical lines are whiskers and they show either the
maximum/minimum values or 1.5 times the interquartile range of the data,
whichever is smaller. Points more than 1.5 times the interquartile range
below (above) the first (third) quartile are defined as outlier and plotted
individually. As seen from Figure~\ref{fig:boxplot}, for each box plot
(across different types of food expenditure) there are large number of
outliers towards the higher values making the distribution positively skewed.
Consequently, the mean food expenditure (which is pushed upward due to the
presence of high values) and covariate effects at the conditional mean is
inadequate for a complete picture. In the literature, studies have used
logarithmic transformation of food expenditure to alleviate this problem of
heteroscedasticity \citep{Liu-et-al-2013}. However, taking a logarithmic
transformation cannot eliminate the non-normality or the heteroscedasticity
problem. Besides, food and nutritional assistance programs (such as
Supplemental Nutrition Assistance Program or SNAP) are typically interested
in the lower tail (i.e., families/households with low food expenditure) to
ensure food security.

The covariates or independent variables utilized in this study (see
Table~\ref{Table:DataSummary}) include age of the head (\emph{Head Age}),
education of the head (\emph{Head Edu}) and the spouse (\emph{Spouse Edu}),
measured as the number of years of schooling and takes value between 0 to 17
(17 represents post graduate level work and above). \emph{Family Size}
represents number of members in a family unit. The variable \emph{Family
Income} indicates the actual value of income including transfer income in the
previous year (negative values representing loss). We use the
inverse-hyperbolic sine (IHS) transformation on income variable because it
adjusts for skewness and retains 0 and negative values
\citep{Friedline-2015,Rahman-Vossmeyer-2019}. The indicator for employment
status of the head (\emph{Head Emp}) equals 1 if the head is employed and 0
otherwise (omitted). The omitted category includes respondents that are
temporarily laid off, looking for work, retired, permanently/temporarily
disabled, keeping house, student and others.

The indicator for gender of the head (\emph{Head Female}) is coded as 1 if
`female' and 0 if `male', while the marital status (of the head) is
categorized into \emph{Married}, \emph{Single} and \emph{Separated}
(omitted). The omitted category (\emph{Separated}) consists of respondents
who are widowed, divorced/annulled or separated. Other variables included in
our study are indicators for homeownership (\emph{Homeowner}) and mortgages
(\emph{Mortgage}). The variable \emph{Homeowner} takes the value 1 if the
respondent is a homeowner and 0 otherwise. Similarly, we have the
\emph{Mortgage} variable which equals 1 if the respondent has a mortgage on
property and 0 otherwise. Race is categorized into \emph{White} and
\emph{Non-White} composed of Blacks, American Indian, Aleut, Eskimo, Asian,
Pacific Islander and Latino. Besides, we have indicators for recession years
and the region in which the family resides. The recession dummy takes the
value 1 for the years 2001, 2007 and 2009 because these were years with some
recession period. Following the US Census Bureau, the region variable is
classified into \emph{Northeast}, \emph{West}, \emph{South} and
\emph{Midwest} (omitted). Including regional indicators help us to look at
differences, if any, in the expenditure behaviour of the families across
regions.

We now look at the movement in average values of the variables for the
sampled period. The average FAH expenditure for a typical family unit is
around \$4590 in 2001, while the FAFH expenditure is around \$1990 for the
same year. The average expenditure on TF is approximately \$6700 in 2001 and
increases to \$8900 in 2015. Not surprisingly, the average expenditure on
FAFH and TF were lower in the year 2009 compared to its respective values in
2007. This shows the adverse effect of the economic crisis on average food
expenditure. The adverse effect seems to persist longer for FAFH expenditure,
as its average value in 2011 is lower compared to 2007.

The average age of the head is around 45 years with a family size of
approximately 3 members in 2001. In the sample, the family units are
predominantly headed by males (about 82\%) with an average of 13.38 years of
schooling in 2001. The average years of schooling of the spouse is lower than
that of the head and stands at 9.5 years in 2001, but increases to
approximately 10 years in 2015. The sample clearly shows the effect of the
Great Recession (December 2007 - June 2009) on the variables \emph{Family
Income} and \emph{Head Emp}. The mean annual family income is approximately
\$77,000 in 2001 and increases to \$99,300 in 2015. However, there is a drop
in average family income for 2011 compared to 2009. The effect of the
economic crisis is much more pronounced on employment status of the head. In
the sample, about 85\% are employed in 2001, which started decreasing in 2005
and stood at 73\% during 2009. However, the lowest percentage employed for
the sample is 62\% in 2015.

A large proportion of the sampled respondents are married (0.68 in 2001) and
remains in the range $0.68-0.71$ throughout the period of our study, while
the proportion of single decreases from 0.15 to 0.09 between 2001 and 2015.
Approximately 75\% of the families own a house in 2001 and this proportion
reaches 81\% in 2007, and remains in that region for subsequent years. The
proportion of respondents having a mortgage on property decrease from 0.59 to
0.53 between 2001 to 2015. Nonetheless, the mortgage percentage was higher
than 0.59 between 2003 to 2009, which is another hallmark of the Great
Recession. On the racial aspect, majority of the sampled families (about
68\%) are White, while the remaining 32\% consists of Blacks and other races;
thus giving a diverse sample for the study. Our sample is also geographically
heterogeneous. Most of the sampled respondents live in the South (38\%),
followed by Midwest (26\%), West (20\%) and Northeast (16\%). This percentage
is stable over the sample period, suggesting little geographic mobility
across regions.


\section{Results}\label{sec:Results}

This section discusses the results for the three types of food expenditure
using the models presented in Section~\ref{sec:Methodology}. In particular,
the results from longitudinal mean regression is presented in
Table~\ref{Table:MeanRegression} and the results from longitudinal quantile
regression is exhibited in Table~\ref{Table:QuantReg}. The posterior
estimates are based on 12,000 MCMC iterations after a burn-in of 3,000
iterations. Trace plots of the MCMC draws, not presented for the sake of
brevity, mimics that of white noise and confirms that the chains have
converged. Moderately diffused priors are utilized for the parameters in both
the models: $\beta \sim N_{k}(0,100*I)$, $\alpha_{i} \sim N_{l}(0,I)$,
$\Sigma^{-1} \sim Wish(5,10 \ast I_{l})$ and $h \sim IG(10/2, 9/2)$. Note
that the definition of $h$ are different in the mean and quantile regression
models. Besides, there are two components of $\alpha_{i}$,
individual-specific intercept and individual-specific coefficient for
inverse-hyperbolic sine transformation of income. With respect to
individual-specific effects, results from Table~\ref{Table:MeanRegression}
and Table~\ref{Table:QuantReg} show that the standard deviations of
$\alpha_{i}$ (i.e., $(\sqrt{\sigma_{11}}, \sqrt{\sigma_{22}}$) are different
for the mean and quantile regression models. As such, a modeling approach
with identical variances should be avoided. We now discuss the results for
the common parameters in all the econometric models.

\begin{table}[!b]
\centering \footnotesize \setlength{\tabcolsep}{4pt}
\setlength{\extrarowheight}{2pt}
\setlength\arrayrulewidth{1pt}
\begin{tabular}{l rrr rrr rrr}
\toprule
&& \multicolumn{2}{c}{\textsc{tf}}
&& \multicolumn{2}{c}{\textsc{fah}}
&& \multicolumn{2}{c}{\textsc{fafh}}\\

\cmidrule{3-4} \cmidrule{6-7} \cmidrule{9-10}
                & &  \textsc{mean} & \textsc{std}
                & &  \textsc{mean} & \textsc{std}
                & &  \textsc{mean} & \textsc{std}\\

\midrule

Intercept           && $-12.53$& $0.84$ &&  $-10.28$ & $ 0.63$    && $-2.12$  & $0.43$ \\

log (HeadAge)       && $ 3.65$ & $0.19$ &&  $ 2.94$  & $ 0.14$    && $ 0.66$  & $0.10$ \\

Head Edu            && $ 0.08$ & $0.02$ &&  $ 0.04$  & $ 0.01$    && $ 0.05$  & $0.01$ \\

Spouse Edu          && $ 0.06$ & $0.01$ &&  $ 0.05$  & $ 0.01$    && $ 0.01$  & $0.01$ \\

Family Size         && $ 0.76$ & $0.03$ &&  $ 0.73$  & $ 0.02$    && $ 0.03$ & $0.02$  \\

IHS Income          && $ 2.79$ & $0.14$ &&  $ 1.37$  & $ 0.10$    && $ 1.35$ & $0.07$  \\

Head Emp(HE)        && $ 0.15$ & $0.08$ &&  $ 0.04$  & $ 0.06$    && $ 0.12$ & $0.05$  \\

Head Female (HF)    && $-1.51$ & $0.20$ &&  $-0.79$  & $ 0.15$    && $-0.71$ & $0.11$  \\

HE$\times$HF        && $ 0.62$ & $0.17$ &&  $ 0.37$  & $ 0.13$    && $ 0.26$ & $0.09$  \\

Married             && $-0.09$ & $0.19$ &&  $-0.01$  & $ 0.14$    && $-0.02$ & $0.10$  \\

Single              && $ 0.96$ & $0.17$ &&  $ 0.44$  & $ 0.12$    && $ 0.43$ & $0.09$  \\

Mort$\times$Home    && $ 0.10$ & $0.07$ &&  $ 0.10$  & $ 0.06$    && $ 0.02$ & $0.04$  \\

Non-White           && $-0.59$ & $0.12$ &&  $-0.35$  & $ 0.09$    && $-0.26$ & $0.06$  \\

Recession           && $-0.26$ & $0.05$ &&  $-0.19$  & $ 0.04$    && $-0.08$ & $0.03$  \\

Northeast           && $ 0.63$ & $0.17$ &&  $ 0.47$  & $ 0.12$    && $ 0.14$ & $0.09$  \\

West                && $ 0.64$ & $0.15$ &&  $ 0.53$  & $ 0.11$    && $ 0.12$ & $0.08$  \\

South               && $ 0.54$ & $0.13$ &&  $ 0.35$  & $ 0.10$    && $ 0.21$ & $0.07$  \\

$h$
                    && $ 0.13$ & $0.01$ &&  $ 0.22$  & $ 0.01$    && $ 0.42$ & $0.01$  \\

$\sigma_{11}^{\frac{1}{2}}$
                    && $ 2.31$ & $0.10$ &&  $ 1.57$  & $ 0.08$    && $ 0.99$ & $0.06$  \\

$\sigma_{22}^{\frac{1}{2}}$
                    && $ 3.07$ & $0.13$ &&  $ 1.87$  & $ 0.10$    && $ 1.66$ & $0.07$  \\

$\rho_{1,2}$
                    && $-0.42$ & $0.05$ &&  $-0.32$  & $ 0.07$    && $-0.28$ & $0.06$  \\

\bottomrule
\multicolumn{10}{l}{\emph{Note}: $h = \sigma^{-2}$ in mean regression.}
\end{tabular}
\caption{Posterior mean (\textsc{mean}) and
standard deviation (\textsc{std}) of the parameters from longitudinal mean regression.}
\label{Table:MeanRegression}
\end{table}

The results from the longitudinal mean regression, presented in
Table~\ref{Table:MeanRegression}, shows that (logarithm of) \emph{Head Age}
positively affects expenditures on TF, FAH and FAFH.  Comparing the
coefficients across categories, we observe that the coefficient for logarithm
of \emph{Head Age} in the FAH equation is much higher (more than 4 times)
than its corresponding value in the FAFH equation. The result agrees with the
intuition that people prefer eating at home as they get older because FAH is
considered to be much healthier. Another argument put forward by
\citet{Liu-et-al-2013} is that social activity reduces with age leading to
lower rise in FAFH expenditure. Other studies that have found a positive
coefficient for \emph{Head Age} include \citet{Redman-1980},
\citet{Nayga-1996}, \citet{Stewart-Yen-2004}, and \citet{Zheng-etal-2019}.
Moving to the results from quantile regression shown in
Table~\ref{Table:QuantReg}, we observe that there is considerable variation
in the coefficients for logarithm of \emph{Head Age}. For example, in the FAH
(FAFH) equation the ratio of coefficients from \emph{Head Age} between
80th-to-20th quantiles is 1.86 (3.58). These differences show considerable
heterogeneity in the effect of \emph{Head Age} on different types of food
expenditure.

The two education variables \emph{Head Edu} and \emph{Spouse Edu} positively
affects TF and FAH expenditures. \citet{Zheng-etal-2019} also finds a
positive effect of head's education on FAH expenditure. For the FAFH
expenditure, only \emph{Head Edu} has a positive effect, but \emph{Spouse
Edu} has no effect (statistically speaking) because the credible interval for
\emph{Spouse Edu} contains zero. This implies that higher educated spouses
(mostly females in our sample) are more knowledgable to understand the
importance of healthy diet and consequently spend more on FAH, but not on
FAFH. Our findings are similar to those reported by \citet{Redman-1980}, and
\citet{Kohara-Kamiya-2016}. The results from quantile regression show
considerable heterogeneity in the covariate effects, but a comparison of the
coefficients for \emph{Head Edu} and \emph{Spouse Edu} seems more
interesting. For FAH expenditure, across quantiles the coefficient for
\emph{Spouse Edu} is always higher than that of \emph{Head Edu} (at the
median the coefficient of \emph{Head Edu} is 0.49 and that of \emph{Spouse
Edu} is 0.51). This implies that spouses (mostly female in our sample) have a
larger positive impact on FAH expenditure across its distribution. In
contrast, for FAFH expenditure the coefficients for \emph{Head Edu} are
higher across quantiles compared to \emph{Spouse Edu}). This implies that an
increase in head's education leads to a higher increase in consumption of
outside food. A possible explanation of such a result is higher involvement
of males in sociable activities \citep{Liu-et-al-2013}.

The variable \emph{Family Size} positively affects TF and FAH expenditures,
but not the FAFH expenditure. The positive effect on FAH is understandable as
larger families tend to eat more at home and less outside, and is consistent
with results reported by \citet{Zheng-etal-2019}. However, the statistically
zero effect on FAFH expenditure is in contrast to those reported in the
literature. While some articles find a positive effect of family size
\citep{Stewart-Yen-2004, Liu-et-al-2013,Zheng-etal-2019}; others have
reported a negative effect on FAFH expenditure
\citep{Redman-1980,Byrne-etal-1996}. The quantile regression results once
again show heterogeneity in covariate effects. For TF expenditure, the
coefficient of \emph{Family Size} is larger at higher quantiles, with the
ratio of 80th-to-20th quantile coefficients at 1.43. For the FAH expenditure,
the coefficient for \emph{Family Size} are similar in size and sign to those
from the TF expenditure equation. Interestingly, \emph{Family Size} has no
impact on FAFH expenditure for lower and middle quantiles.

\begin{landscape}
\begin{table}[!t]
\centering \footnotesize \setlength{\tabcolsep}{3pt}
\setlength{\extrarowheight}{2pt}
\setlength\arrayrulewidth{1pt}
\begin{tabular}{l rrr rrr rrr rrr rrr rrr rrr rrr rrr}
\toprule
&&\multicolumn{8}{c}{\textsc{tf}} && \multicolumn{8}{c}{\textsc{fah}}
&& \multicolumn{8}{c}{\textsc{fafh}}\\
\cmidrule{3-10} \cmidrule{12-19} \cmidrule{21-28}
&& \multicolumn{2}{c}{\textsc{20$^{th}$ }}
&& \multicolumn{2}{c}{\textsc{50$^{th}$ }}
&& \multicolumn{2}{c}{\textsc{80$^{th}$ }}
&& \multicolumn{2}{c}{\textsc{20$^{th}$ }}
&& \multicolumn{2}{c}{\textsc{50$^{th}$ }}
&& \multicolumn{2}{c}{\textsc{80$^{th}$ }}
&& \multicolumn{2}{c}{\textsc{20$^{th}$ }}
&& \multicolumn{2}{c}{\textsc{50$^{th}$ }}
&& \multicolumn{2}{c}{\textsc{80$^{th}$ }} \\

\cmidrule{3-4} \cmidrule{6-7} \cmidrule{9-10} \cmidrule{12-13} \cmidrule{15-16}
\cmidrule{18-19} \cmidrule{21-22} \cmidrule{24-25} \cmidrule{27-28}

                & &  \textsc{mean} & \textsc{std}
                & &  \textsc{mean} & \textsc{std}
                & &  \textsc{mean} & \textsc{std}
                & &  \textsc{mean} & \textsc{std}
                & &  \textsc{mean} & \textsc{std}
                & &  \textsc{mean} & \textsc{std}
                & &  \textsc{mean} & \textsc{std}
                & &  \textsc{mean} & \textsc{std}
                & &  \textsc{mean} & \textsc{std}\\
\midrule
Intercept           && $-9.22$ & $0.61$ && $-11.72$  & $ 0.70$    && $-14.94$ & $0.86$
                    && $-7.83$ & $0.44$ && $-10.24$  & $ 0.52$    && $-13.15$ & $0.66$
                    && $-0.69$ & $0.22$ &&  $-1.12$  & $ 0.30$    && $ -1.48$ & $0.43$\\

log(Head Age)       && $ 2.56$  & $0.14$ && $ 3.41$  & $ 0.16$    && $ 4.68$  & $0.20$
                    && $ 2.13$  & $0.10$ && $ 2.88$  & $ 0.12$    && $ 3.97$  & $0.15$
                    && $ 0.19$  & $0.05$ && $ 0.40$  & $ 0.07$    && $ 0.68$  & $0.10$\\

Head Edu            && $ 0.06$ & $0.01$ &&  $ 0.08$  & $ 0.02$    && $ 0.06$  & $0.02$
                    && $ 0.03$ & $0.01$ &&  $ 0.05$  & $ 0.01$    && $ 0.03$  & $0.02$
                    && $ 0.02$ & $0.01$ &&  $ 0.03$  & $ 0.01$    && $ 0.04$  & $0.01$\\

Spouse Edu          && $ 0.05$ & $0.01$ &&  $ 0.06$  & $ 0.01$    && $ 0.06$  & $0.01$
                    && $ 0.04$ & $0.01$ &&  $ 0.05$  & $ 0.01$    && $ 0.06$  & $0.01$
                    && $ 0.01$ & $0.01$ &&  $ 0.01$  & $ 0.01$    && $ 0.01$  & $0.01$\\

Family Size         && $ 0.58$ & $0.02$ &&  $ 0.70$  & $ 0.02$    && $ 0.83$ & $0.03$
                    && $ 0.55$ & $0.02$ &&  $ 0.67$  & $ 0.02$    && $ 0.80$ & $0.02$
                    && $-0.01$ & $0.01$ &&  $ 0.01$  & $ 0.01$    && $ 0.04$ & $0.01$\\

IHS Income          && $ 2.33$ & $0.10$ &&  $ 2.71$  & $ 0.12$    && $ 3.22$ & $0.16$
                    && $ 1.06$ & $0.07$ &&  $ 1.24$  & $ 0.08$    && $ 1.50$ & $0.11$
                    && $ 1.00$ & $0.04$ &&  $ 1.33$  & $ 0.06$    && $ 1.71$ & $0.08$\\

Head Emp(HE)        && $ 0.21$ & $0.06$ &&  $ 0.16$  & $ 0.07$    && $ 0.08$ & $0.08$
                    && $ 0.05$ & $0.04$ &&  $ 0.06$  & $ 0.05$    && $ 0.03$ & $0.06$
                    && $ 0.06$ & $0.02$ &&  $ 0.08$  & $ 0.03$    && $ 0.08$ & $0.04$\\

Head Female (HF)    && $-0.99$ & $0.15$ &&  $-1.36$  & $ 0.17$    && $-1.80$ & $0.23$
                    && $-0.50$ & $0.11$ &&  $-0.74$  & $ 0.13$    && $-1.04$ & $0.18$
                    && $-0.29$ & $0.05$ &&  $-0.53$  & $ 0.07$    && $-0.83$ & $0.12$\\

HE$\times$HF        && $ 0.35$ & $0.11$ &&  $ 0.45$  & $ 0.13$    && $ 0.47$ & $0.15$
                    && $ 0.24$ & $0.08$ &&  $ 0.26$  & $ 0.09$    && $ 0.22$ & $0.12$
                    && $ 0.15$ & $0.04$ &&  $ 0.18$  & $ 0.06$    && $ 0.14$ & $0.08$\\

Married             && $ 0.12$ & $0.14$ &&  $ 0.02$  & $ 0.16$    && $-0.11$ & $0.20$
                    && $ 0.11$ & $0.10$ &&  $ 0.01$  & $ 0.12$    && $-0.06$ & $0.15$
                    && $ 0.04$ & $0.05$ &&  $-0.02$  & $ 0.07$    && $ 0.01$ & $0.10$\\

Single              && $ 0.73$ & $0.12$ &&  $ 0.96$  & $ 0.15$    && $ 1.17$ & $0.20$
                    && $ 0.33$ & $0.09$ &&  $ 0.49$  & $ 0.11$    && $ 0.56$ & $0.14$
                    && $ 0.20$ & $0.04$ &&  $ 0.29$  & $ 0.06$    && $ 0.47$ & $0.10$\\

Mort$\times$Home    && $ 0.02$ & $0.05$ &&  $ 0.06$  & $ 0.06$    && $ 0.09$ & $0.07$
                    && $ 0.06$ & $0.04$ &&  $ 0.07$  & $ 0.04$    && $ 0.06$ & $0.05$
                    && $ 0.01$ & $0.02$ &&  $ 0.01$  & $ 0.03$    && $ 0.01$ & $0.03$\\

Non-White           && $-0.61$ & $0.09$ &&  $-0.50$  & $ 0.11$    && $-0.45$ & $0.14$
                    && $-0.38$ & $0.06$ &&  $-0.30$  & $ 0.08$    && $-0.20$ & $0.11$
                    && $-0.19$ & $0.03$ &&  $-0.21$  & $ 0.05$    && $-0.24$ & $0.07$\\

Recession           && $-0.12$ & $0.03$ &&  $-0.15$  & $ 0.03$    && $-0.15$ & $0.04$
                    && $-0.06$ & $0.02$ &&  $-0.08$  & $ 0.03$    && $-0.07$ & $0.03$
                    && $-0.04$ & $0.01$ &&  $-0.04$  & $ 0.02$    && $-0.03$ & $0.02$\\

Northeast           && $ 0.61$ & $0.12$ &&  $ 0.69$  & $ 0.15$    && $ 0.66$ & $0.20$
                    && $ 0.41$ & $0.09$ &&  $ 0.44$  & $ 0.11$    && $ 0.52$ & $0.15$
                    && $ 0.04$ & $0.05$ &&  $ 0.12$  & $ 0.07$    && $ 0.25$ & $0.10$\\

West                && $ 0.38$ & $0.12$ &&  $ 0.56$  & $ 0.14$    && $ 0.66$ & $0.18$
                    && $ 0.29$ & $0.08$ &&  $ 0.43$  & $ 0.10$    && $ 0.62$ & $0.13$
                    && $ 0.03$ & $0.04$ &&  $ 0.09$  & $ 0.06$    && $ 0.15$ & $0.09$ \\

South               && $ 0.35$ & $0.10$ &&  $ 0.48$  & $ 0.12$    && $ 0.68$ & $0.15$
                    && $ 0.20$ & $0.07$ &&  $ 0.26$  & $ 0.09$    && $ 0.41$ & $0.11$
                    && $ 0.11$ & $0.04$ &&  $ 0.17$  & $ 0.05$    && $ 0.26$ & $0.08$\\

$h$
                    && $ 1.76$ & $0.02$ &&  $ 1.10$  & $ 0.01$    && $ 1.49$ & $0.01$
                    && $ 2.31$ & $0.02$ &&  $ 1.44$  & $ 0.01$    && $ 1.93$ & $0.02$
                    && $ 3.64$ & $0.03$ &&  $ 2.18$  & $ 0.02$    && $ 2.84$ & $0.02$\\

$\sigma_{11}^{\frac{1}{2}}$
                    && $ 1.72$ & $0.08$ &&  $ 2.09$  & $ 0.09$    && $ 3.47$ & $0.11$
                    && $ 1.20$ & $0.05$ &&  $ 1.42$  & $ 0.06$    && $ 2.53$ & $0.08$
                    && $ 0.38$ & $0.04$ &&  $ 0.72$  & $ 0.05$    && $ 1.63$ & $0.06$\\

$\sigma_{22}^{\frac{1}{2}}$
                    && $ 2.45$ & $0.11$ &&  $ 2.81$  & $ 0.12$    && $ 4.52$ & $0.15$
                    && $ 1.48$ & $0.08$ &&  $ 1.55$  & $ 0.10$    && $ 2.87$ & $0.12$
                    && $ 0.95$ & $0.04$ &&  $ 1.40$  & $ 0.06$    && $ 2.44$ & $0.08$\\
$\rho_{1,2}$
                    && $-0.45$ & $0.05$ &&  $-0.36$  & $ 0.05$    && $-0.56$ & $0.03$
                    && $-0.31$ & $0.06$ &&  $-0.14$  & $ 0.08$    && $-0.52$ & $0.04$
                    && $ 0.14$ & $0.14$ &&  $-0.06$  & $ 0.09$    && $-0.43$ & $0.04$\\
\bottomrule
\multicolumn{27}{l}{\emph{Note}: $h = \sigma^{-1}$ in quantile regression.}
\end{tabular}
\caption{Posterior mean (\textsc{mean}) and
standard deviation (\textsc{std}) of the parameters from longitudinal quantile
regression.}
\label{Table:QuantReg}
\end{table}
\end{landscape}

Total family income is perhaps the most decisive variable that steers food
expenditure. We use the IHS transformation of family income for reasons
mentioned earlier \citep[see also][]{Friedline-2015,Rahman-Vossmeyer-2019}.
As seen in Table~\ref{Table:MeanRegression}, the transformed income variable
positively affects expenditures on TF, FAH and FAFH. The intuition is clear,
increase in income translates to increase in food expenditures of all types.
This result finds support in several other works such as \citet{Redman-1980},
\citet{Lee-Brown-1986}, \citet{Nayga-1996}, \citet{ZiolGuest-etal-2006}, and
\citet{Liu-et-al-2013}. Results from quantile regression show considerable
heterogeneity in covariate effects with higher quantiles showing a larger
impact of income on food expenditure. The ratio of 80th-to-20th quantile
coefficients for \emph{IHS Income} in the TF, FAH and FAFH equations are
1.38, 1.41 and 1.71, respectively.

The next three variables in Table~\ref{Table:MeanRegression} are indicator
variable for head's employment (\emph{Head Emp}), indicator variable for
female head (\emph{Head Female}), and interaction of the two indicators.
Head's employment has a positive effect on TF and FAFH expenditures, but
statistically has no effect on FAH expenditure. These findings are similar to
those in \citet{Aguiar-Hurst-2005}, \citet{Huang-etal-2015}, and
\citet{Antelo-etal-2017}. The indicator for \emph{Head Female} is negative
for all categories, which suggests that female headed families tend to spend
less on overall and each category of food. This can be attributed to two
factors, females are better at managing family expenditure and an empowered
woman better understands the importance of nutritious food and thus reduces
FAFH expenditure. The interaction term (\emph{Head Emp} $\times$ \emph{Head
Female}) in all three regressions are positive, which implies that an
employed female head spends more on overall and each category of food
purchase. Results from quantile regression, presented in
Table~\ref{Table:QuantReg}, once again reveal heterogeneity in the covariate
effect of the three indicator variables. Heads's employment positively
affects TF expenditure at lower and middle quantiles, but not at upper
quantiles. There is no effect on FAH expenditure and a positive effect on
FAFH expenditure across quantiles. \emph{Head Female} has a negative effect
on overall and each category of food expenditure, and the negative effect
increases at upper quantiles. The interaction term shows a positive effect on
TF expenditure across quantiles, but a positive effect on FAH and FAFH
expenditures only at lower and middle quantiles. Hence, at higher levels of
FAH and FAFH expenditures the employment of female head does not play an
important role.

The impact of marital status on food expenditure is examined through the two
indicator variables, \emph{Married} and \emph{Single}. The base or omitted
category is \emph{Separated}, explained in Section~\ref{sec:Data}. As seen
from Table~\ref{Table:MeanRegression}, the coefficient for \emph{Married} is
not statistically different from zero. So, being married has statistically no
effect on food expenditure relative to the omitted category,
\emph{Separated}. However, being single has a positive effect on overall food
expenditure and across categories. Our findings are consistent with results
reported by \citet{Stewart-Yen-2004} and \citet{Liu-et-al-2013}, but
contradictory to those by \citet{Byrne-etal-1996} and
\citet{Zheng-etal-2019}. The results from quantile regression reinforces the
findings from the mean regression. Across quantiles, being married has no
effect on food expenditure as compared to the omitted category. On the other
hand, being single has a positive effect on food expenditure and are
increasing with quantiles. The ratio of 80th-to-20th quantile coefficients
for TF, FAH and FAFH expenditures are 1.60, 1.70, and 2.35, respectively.

Homeowners having mortgages are resource constrained and have a lower cash
flow for a given income. This may negatively affect food expenditure,
particularly, FAFH expenditure. To explore this hypothesis, we include an
indicator variable for homeowners having mortgages into our regression
equations. Results from Table~\ref{Table:MeanRegression} and
Table~\ref{Table:QuantReg} show that families with mortgages have
statistically no effect on food expenditure. Our results are opposite to
those by \citet{Nayga-1996}, where he finds that homeowners with mortgages do
spend more on food prepared at home and FAFH, but not on prepared foods
(e.g., frozen meals and prepared salads). Similarly, \citet{Liu-et-al-2013}
find that homeowners who are married (with and without children) have higher
probability of different types of FAFH expenditures (e.g., full-service
dining, fast-food and other facilities), but for single-person homeowners
this is true only for full-service dining.

Variations in food expenditure have often been linked to racial disparity. To
investigate this conjecture, we include an indicator variable for
\emph{Non-White}, keeping \emph{White} as the base or omitted category.
Results from mean regression, presented in Table~\ref{Table:MeanRegression},
exhibit that \emph{Non-White} tends to have lower expenditure on overall
food, as well as FAH and FAFH expenditures. Our findings are consistent with
\citet{Nayga-1996} who finds that white households are likely to spend more
on FAH and FAFH. Similarly, \citet{Lee-Brown-1986} report that Non-White are
less likely to eat away from home. Our results are also in agreement with
findings from other previous works such as \citet{Redman-1980},
\citet{Stewart-Yen-2004}, and \citet{Liu-et-al-2013}. Another reason for the
negative coefficient, as noted by \citet{Byrne-etal-1996} is due to
non-availability of ethnic foods at local restaurants. The results from
quantile regression, shown in Table~\ref{Table:QuantReg}, largely agree with
the finding from mean regression. \emph{Non-White} have lower TF expenditure
compared to \emph{White}. Moreover, the impact is larger at lower quantiles
and decrease as we move to upper quantiles. For, FAH expenditure, the
\emph{Non-White} have lower expenditures only at the lower and middle
quantiles, but not at upper quantiles. In contrast, FAFH expenditure for
\emph{Non-White} is lower across quantiles and the negative impact increases
with increasing quantiles.

Most expenditures, including consumption, typically decline during times of
recession. To explore the negative effect on food expenditure, if any, we
include an indicator for recession years (2001, 2007 and 2009) into our
regression. Results from mean regression show that the coefficient for
\emph{Recession} is negative for all types of food expenditure, which implies
that expected food expenditure (overall and category wise) declined during
the recession years. As reported in Table~\ref{Table:MeanRegression}, average
TF, FAH and FAFH expenditures declined by \$257, \$190 and \$75,
respectively. Our findings are supported by \citet{Griffith-etal-2013}, where
they report decline in expenditure for food items for British households
during and post the Great Recession. Similarly, \citet{Antelo-etal-2017} also
find that food expenditure for Spanish households declined during the crisis
period (i.e., 2008$-$2014) in Spain. Moving to quantile regression, we find
that the quantile results reinforces the findings from mean regression. Both
TF and FAH expenditures declined across quantiles during the recession years,
and the effect is more or less uniform across the considered quantiles. For
FAFH expenditure, we observe a decline only at lower and middle quantiles,
but not at the upper quantile. So, families whose expenditure on FAFH is high
are not affected by recession years.

Lastly, we include regional indicators to examine geographical differences in
food expenditure. These differences may be due to varying levels of
urbanization, climatic conditions and diverse food culture. We include
indicators for \emph{Northeast}, \emph{West} and \emph{South} into our
regression equations. \emph{Midwest} is used as the base or omitted category.
Our regional classification follows the definition of the US Census Bureau.
Results from mean regressions (see Table~\ref{Table:MeanRegression}) reveal
that an average family living in \emph{South} (relative to \emph{Midwest})
have higher TF, FAH and FAFH expenditures. However, for families living in
the \emph{Northeast} and \emph{West}, the average expenditure is more on TF
and FAH but not on FAFH. Other studies, such as \citet{Lee-Brown-1986},
\citet{Nayga-1996}, \citet{Byrne-etal-1996} and \citet{Liu-et-al-2013} also
find disparity in regional food expenditures. Moving to results from quantile
regression (see Table~\ref{Table:QuantReg}), we see that for TF and FAH
expenditures, all the quantile coefficients for \emph{Northeast}, \emph{West}
and \emph{South} are positive and increase with quantiles. This suggests
families living in the three regions have higher quantile expenditures
(compared to those living in \emph{Midwest}) and the differential impact
increases at higher quantiles. For FAFH expenditure, only \emph{South} and
\emph{Northeast} (at the upper quantiles only) have a positive effect on FAFH
expenditure.

In summary, the results from quantile regression reveal considerable
heterogeneity in covariate effects which cannot be uncovered from mean
regression. The additional information from quantile regression may be useful
for policy making in the government or business, such as aiming sections of
the population for welfare schemes or running campaigns to promote business.

\subsection{Heterogeneity Bias}

Unobserved heterogeneity is a large component of food expenditure and we
control for this in our (mean and quantile) regression models with
individual-specific parameters in the intercept and income. To demonstrate
the heterogeneity bias and poorer model fit that can occur, we estimate the
quantile models without including the individual-specific effects (i.e.,
without including the conditional dependence between observations across time
for the same family unit). This model can be estimated as a special case of
Algorithm~\ref{algo2}, by eliminating Step~1(a) and Step~(3), and removing
$(\alpha_{i}, \Sigma^{-1})$ from the conditional posteriors of the remaining
parameters.

The results from the longitudinal quantile models without the
individual-specific effects are presented in Table~\ref{Table:QuantRegNoRE}
and they differ widely compared to those of Table~\ref{Table:QuantReg}, which
presents the results from longitudinal quantile regression with
individual-specific effects. For example, the coefficients for \emph{Head
Age}, \emph{IHS Income}, \emph{Head Female}, and \emph{Recession} are
noticeably different in the two models across quantiles and types of food
expenditure. Again, there are variables whose coefficients either become
statistically equivalent to or different from zero when the
individual-specific parameters are excluded. In the former category, we have
the coefficient for \emph{Spouse Edu} at middle and upper quantile for total
food expenditure. In the latter category, we have the coefficient for
homeowners with mortgages (\emph{Mort $\times$ Home}) at lower quantiles for
expenditures on total food and food at home.

\begin{landscape}
\begin{table}
\centering \footnotesize \setlength{\tabcolsep}{3pt}
\setlength{\extrarowheight}{2pt}
\setlength\arrayrulewidth{1pt}
\begin{tabular}{l rrr rrr rrr rrr rrr rrr rrr rrr rrr}
\toprule
&&\multicolumn{8}{c}{\textsc{tf}} && \multicolumn{8}{c}{\textsc{fah}}
&& \multicolumn{8}{c}{\textsc{fafh}}\\
\cmidrule{3-10} \cmidrule{12-19} \cmidrule{21-28}
&& \multicolumn{2}{c}{\textsc{20$^{th}$ }}
&& \multicolumn{2}{c}{\textsc{50$^{th}$ }}
&& \multicolumn{2}{c}{\textsc{80$^{th}$ }}
&& \multicolumn{2}{c}{\textsc{20$^{th}$ }}
&& \multicolumn{2}{c}{\textsc{50$^{th}$ }}
&& \multicolumn{2}{c}{\textsc{80$^{th}$ }}
&& \multicolumn{2}{c}{\textsc{20$^{th}$ }}
&& \multicolumn{2}{c}{\textsc{50$^{th}$ }}
&& \multicolumn{2}{c}{\textsc{80$^{th}$ }} \\

\cmidrule{3-4} \cmidrule{6-7} \cmidrule{9-10} \cmidrule{12-13} \cmidrule{15-16}
\cmidrule{18-19} \cmidrule{21-22} \cmidrule{24-25} \cmidrule{27-28}

                & &  \textsc{mean} & \textsc{std}
                & &  \textsc{mean} & \textsc{std}
                & &  \textsc{mean} & \textsc{std}
                & &  \textsc{mean} & \textsc{std}
                & &  \textsc{mean} & \textsc{std}
                & &  \textsc{mean} & \textsc{std}
                & &  \textsc{mean} & \textsc{std}
                & &  \textsc{mean} & \textsc{std}
                & &  \textsc{mean} & \textsc{std}\\

\midrule

Intercept           && $-0.80$ & $0.39$ && $ -1.26$  & $ 0.52$    && $ -1.70$ & $0.66$
                    && $-1.77$ & $0.29$ && $ -2.75$  & $ 0.36$    && $ -3.46$ & $0.45$
                    && $ 0.86$ & $0.16$ && $  1.02$  & $ 0.25$    && $  0.35$ & $0.39$\\

log(Head Age)       && $ 0.46$  & $0.09$ && $ 0.70$  & $ 0.12$    && $ 1.12$  & $0.15$
                    && $ 0.58$  & $0.06$ && $ 0.93$  & $ 0.08$    && $ 1.24$  & $0.10$
                    && $-0.23$  & $0.04$ && $-0.16$  & $ 0.05$    && $ 0.18$  & $0.08$\\

Head Edu            && $ 0.01$ & $0.01$ &&  $ 0.03$  & $ 0.01$    && $ 0.04$  & $0.01$
                    && $-0.01$ & $0.01$ &&  $ 0.02$  & $ 0.01$    && $ 0.04$  & $0.01$
                    && $ 0.01$ & $0.00$ &&  $ 0.02$  & $ 0.01$    && $ 0.02$  & $0.01$\\

Spouse Edu          && $ 0.05$ & $0.01$ &&  $ 0.01$  & $ 0.01$    && $-0.01$  & $0.01$
                    && $ 0.05$ & $0.01$ &&  $ 0.04$  & $ 0.01$    && $ 0.01$  & $0.01$
                    && $-0.01$ & $0.01$ &&  $-0.01$  & $ 0.00$    && $-0.02$  & $0.01$\\

Family Size         && $ 0.52$ & $0.02$ &&  $ 0.80$  & $ 0.02$    && $ 1.14$ & $0.03$
                    && $ 0.54$ & $0.01$ &&  $ 0.78$  & $ 0.02$    && $ 1.12$ & $0.02$
                    && $-0.02$ & $0.01$ &&  $-0.01$  & $ 0.01$    && $ 0.06$ & $0.01$\\

IHS Income          && $ 2.47$ & $0.06$ &&  $ 3.71$  & $ 0.08$    && $ 5.22$ & $0.09$
                    && $ 1.01$ & $0.04$ &&  $ 1.43$  & $ 0.06$    && $ 2.09$ & $0.07$
                    && $ 0.99$ & $0.03$ &&  $ 1.81$  & $ 0.04$    && $ 3.11$ & $0.05$\\

Head Emp(HE)        && $ 0.23$ & $0.06$ &&  $ 0.11$  & $ 0.07$    && $-0.12$ & $0.09$
                    && $ 0.07$ & $0.04$ &&  $ 0.15$  & $ 0.05$    && $ 0.05$ & $0.07$
                    && $ 0.09$ & $0.02$ &&  $ 0.06$  & $ 0.03$    && $-0.01$ & $0.05$\\

Head Female (HF)    && $-0.55$ & $0.08$ &&  $-0.81$  & $ 0.11$    && $-1.41$ & $0.15$
                    && $-0.18$ & $0.06$ &&  $-0.25$  & $ 0.08$    && $-0.43$ & $0.11$
                    && $-0.06$ & $0.03$ &&  $-0.40$  & $ 0.05$    && $-0.86$ & $0.08$\\

HE$\times$HF        && $ 0.42$ & $0.09$ &&  $ 0.51$  & $ 0.12$    && $ 0.68$ & $0.15$
                    && $ 0.28$ & $0.07$ &&  $ 0.21$  & $ 0.09$    && $ 0.19$ & $0.11$
                    && $ 0.01$ & $0.03$ &&  $ 0.18$  & $ 0.05$    && $ 0.27$ & $0.08$\\

Married             && $-0.01$ & $0.12$ &&  $ 0.23$  & $ 0.14$    && $ 0.01$ & $0.16$
                    && $-0.04$ & $0.08$ &&  $ 0.09$  & $ 0.10$    && $ 0.45$ & $0.11$
                    && $ 0.03$ & $0.04$ &&  $-0.01$  & $ 0.07$    && $-0.18$ & $0.09$\\

Single              && $ 0.18$ & $0.06$ &&  $ 0.36$  & $ 0.09$    && $ 0.57$ & $0.12$
                    && $ 0.06$ & $0.05$ &&  $ 0.12$  & $ 0.06$    && $ 0.15$ & $0.08$
                    && $ 0.10$ & $0.03$ &&  $ 0.16$  & $ 0.04$    && $ 0.38$ & $0.06$\\

Mort$\times$Home    && $ 0.11$ & $0.04$ &&  $ 0.02$  & $ 0.05$    && $ 0.03$ & $0.07$
                    && $ 0.14$ & $0.03$ &&  $ 0.03$  & $ 0.04$    && $ 0.01$ & $0.05$
                    && $ 0.01$ & $0.02$ &&  $-0.06$  & $ 0.02$    && $ 0.02$ & $0.04$\\

Non-White           && $-0.67$ & $0.04$ &&  $-0.80$  & $ 0.06$    && $-0.87$ & $0.08$
                    && $-0.48$ & $0.03$ &&  $-0.51$  & $ 0.04$    && $-0.50$ & $0.05$
                    && $-0.11$ & $0.02$ &&  $-0.22$  & $ 0.03$    && $-0.41$ & $0.04$\\

Recession           && $-0.23$ & $0.04$ &&  $-0.36$  & $ 0.05$    && $-0.51$ & $0.06$
                    && $-0.17$ & $0.03$ &&  $-0.20$  & $ 0.03$    && $-0.51$ & $0.04$
                    && $-0.05$ & $0.02$ &&  $-0.07$  & $ 0.02$    && $-0.08$ & $0.03$\\

Northeast           && $ 0.51$ & $0.06$ &&  $ 0.88$  & $ 0.08$    && $ 1.41$ & $0.10$
                    && $ 0.52$ & $0.04$ &&  $ 0.61$  & $ 0.05$    && $ 0.97$ & $0.07$
                    && $-0.01$ & $0.03$ &&  $ 0.12$  & $ 0.04$    && $ 0.32$ & $0.05$\\

West                && $ 0.36$ & $0.05$ &&  $ 0.46$  & $ 0.07$    && $ 0.79$ & $0.09$
                    && $ 0.40$ & $0.04$ &&  $ 0.49$  & $ 0.05$    && $ 0.70$ & $0.06$
                    && $-0.01$ & $0.02$ &&  $ 0.03$  & $ 0.03$    && $ 0.06$ & $0.05$ \\

South               && $ 0.38$ & $0.05$ &&  $ 0.49$  & $ 0.06$    && $ 0.80$ & $0.08$
                    && $ 0.23$ & $0.03$ &&  $ 0.25$  & $ 0.04$    && $ 0.49$ & $0.05$
                    && $ 0.10$ & $0.02$ &&  $ 0.23$  & $ 0.03$    && $ 0.30$ & $0.04$\\

$h$                 && $ 1.26$ & $0.01$ &&  $ 0.76$  & $ 0.01$    && $ 0.90$ & $0.01$
                    && $ 1.71$ & $0.01$ &&  $ 1.04$  & $ 0.01$    && $ 1.24$ & $0.01$
                    && $ 2.70$ & $0.02$ &&  $ 1.48$  & $ 0.01$    && $ 1.62$ & $0.01$\\
\bottomrule
\multicolumn{27}{l}{\emph{Note}: $h = \sigma^{-1}$ in quantile regression.}
\end{tabular}
\caption{Posterior mean (\textsc{mean}) and
standard deviation (\textsc{std}) of the parameters from longitudinal quantile
regression without random-effects.}
\label{Table:QuantRegNoRE}
\end{table}
\end{landscape}

\begin{sloppypar}
To highlight the importance of modeling the individual-specific effects (or
random-effects), we compare model fitting at the considered quantiles using
the conditional log-likelihood, conditional Akaike Information Criterion
(cAIC) and conditional Bayesian Information Criterion (cBIC). The calculation
of cAIC and cBIC are proposed and explained in \citet{Greven-Kneib-2010} and
\citet{Delattre-etal-2014}, respectively. These model comparison measures are
presented in Table~\ref{Table:ModelComp}. The table clearly shows that across
quantiles, the value of the conditional log-likelihood is higher and those of
cAIC and cBIC are lower for each longitudinal quantile regression when
individual-specific effects are included. Consequently, there is a strong
evidence for modeling unobserved heterogeneity and ignoring it can lead to
poor model fitting.
\end{sloppypar}

\begin{table}[!t]
\centering \footnotesize \setlength{\tabcolsep}{5pt} \setlength{\extrarowheight}{1.5pt}
\setlength\arrayrulewidth{1pt}
\begin{tabular}{c rrr rrr rrr r  }
\toprule
& & \multicolumn{2}{c}{\textsc{20th quantile}} & & \multicolumn{2}{c}{\textsc{50th quantile}}
& & \multicolumn{2}{c}{\textsc{80th quantile}} \\
\cmidrule{3-4} \cmidrule{6-7}  \cmidrule{9-10}
         & &  with \textsc{RE} & w/o \textsc{RE}
         & &  with \textsc{RE} & w/o \textsc{RE}
         & &  with \textsc{RE} & w/o \textsc{RE}  &  \\
\midrule
    & & \multicolumn{8}{c}{TF Expenditure} &  \\
\midrule
$\log$-L        & & $ -38282$  & $-45188$  & & $ -38701$ & $-46268$
                & & $ -40996$  & $-51031$  &   \\
cAIC            & & $  76601$  & $ 90411$  & & $  77439$ & $ 92573$
                & & $  82029$  & $102099$  &   \\
cBIC            & & $  76802$  & $ 90551$  & & $  77640$ & $ 92713$
                & & $  82230$  & $102239$  &   \\
\midrule
    & & \multicolumn{8}{c}{FAH Expenditure} &  \\
\midrule
$\log$-L        & & $ -33577$  & $-39932$  & & $ -34054$ & $-40773$
                & & $ -36617$  & $-45538$  &   \\
cAIC            & & $  67189$  & $ 79901$  & & $  68143$ & $ 81582$
                & & $  73271$  & $ 91111$  &   \\
cBIC            & & $  67390$  & $ 80041$  & & $  68344$ & $ 81722$
                & & $  73472$  & $ 91251$  &   \\
\midrule
    & & \multicolumn{8}{c}{FAFH Expenditure} &  \\
\midrule
$\log$-L        & & $ -25753$  & $-31996$  & & $ -26797$ & $-34630$
                & & $ -29883$  & $-40896$  &   \\
cAIC            & & $  51542$  & $ 64028$  & & $  53630$ & $ 69295$
                & & $  59802$  & $ 81828$  &   \\
cBIC            & & $  51743$  & $ 64168$  & & $  53832$ & $ 69435$
                & & $  60003$  & $ 81968$  &   \\
\bottomrule
\end{tabular}
\caption{Model comparison between the longitudinal quantile
regression with random-effects (with RE) and without random-effects (w/o RE).
The log-likelihood ($\log$-L), conditional Akaike Information Criterion (cAIC) and
conditional Bayesian Information Criterion (cBIC) are evaluated at the posterior mean
of the parameters.}\label{Table:ModelComp}
\end{table}

\section{Conclusion}\label{sec:Conclusion}

This article studies the relationship between different types of food
expenditures (total food, food at home, and food away from home) and a host
of economic, geographic, and demographic factors using data from the Panel
Study of Income Dynamics for the period 2001$-$2015. Food expenditures are
typically right skewed and thus covariate effects are likely to be
heterogeneous across the conditional distribution of the response variable.
Besides, unobserved heterogeneity is a large component of food expenditure.
To explore these considerations, we study food expenditure within a
longitudinal quantile framework that models dependence between the
observations across time for the same family units. Results point to several
important aspects including the presence of heterogeneity in the covariate
effects. For example, we find that there are notable differences in the food
expenditure behavior (of all types) between male and female headed
households, expenditures on food away from home by employed female heads are
heterogeneous across quantiles, and food expenditures (of all types) decrease
during times of economic crisis and varies with quantiles. Besides, the paper
provides strong empirical evidence that not considering unobserved
heterogeneity can lead to heterogeneity bias and poor model fitting.

While our paper emphasizes the modeling of heterogeneity in food expenditure,
the findings reported also provide greater insights on total food expenditure
and expenditures on food at home and food away from home, which may be of
special interest to policy makers in the health and business sectors. For
example, we find that spouse education has a positive effect across the
distribution of food at home expenditure. Therefore, policy makers may
provide higher incentives to female education in order to achieve better
health outcomes in the country. Similarly, we find that being single or
employed female heads have a positive effect on the distribution of food away
from home expenditure. Consequently, restaurant and fast food chains may run
campaigns targeting these specific groups to increase their sales. The above
discussion and other findings reported in the paper may be utilized to better
formulate policies and business decisions.


\clearpage \pagebreak
\pdfbookmark[1]{References}{unnumbered}


\bibliography{BibFoodExp}
\bibliographystyle{jasa}

\end{document}